\documentclass[printer]{aa}
\usepackage{natbib}
\usepackage{graphicx}
\begin{document}
\title{ Star count analysis of the interstellar matter
in the region of L1251}
\author{L.G.~Bal\'azs\inst{1} \and P.~\'Abrah\'am\inst{1}
\and M.~Kun\inst{1} \and J.~Kelemen\inst{1} \and
L.V.~T\'oth\inst{2,}\inst{3}} \institute{Konkoly Observatory of
the Hungarian Academy of Sciences, P.O.~Box~67, H-1525 Budapest,
Hungary \and Max-Planck-Institut f\"ur Astronomie (MPIA),
K\"onigstuhl 17, D-69117 Heidelberg, Germany \and E\"otv\"os
Lor\'and University of Sciences, Dept. of Astronomy, P\'azm\'any
P\'eter s\'et\'any 1/A, H-1117 Budapest, Hungary }
\date{Received date; accepted date}
\authorrunning{Bal\'azs et al.}
\titlerunning{Star count analysis ...}
   \abstract{
We studied the ISM distribution in and around the star forming
cloud L1251 with optical star counts. A careful  calculation  with
a maximum likelihood based statistical approach resulted in {\it
B, V, R, I} extinction distributions from the star count maps. A
distance of $330\pm 30$pc was derived. The extinction maps
revealed an elongated dense cloud with a bow shock at its eastern
side. We estimated a Mach number of $M\approx2$ for the bow shock.
A variation of the apparent dust properties is detected, i.e. the
$R_V=A_V/E_{B-V}$ total to selective extinction ratio varies from
3 to 5.5, peaking at the densest part of L1251. The spatial
structure of the head of L1251 is well modelled with a
Schuster-sphere (i.e. n=5 polytropic sphere). The observed radial
distribution of mass fits  the model with  high accuracy out to
2.5pc distance from the assumed center. Unexpectedly, the
distribution of $NH_3$ 1.3 cm line widths is also well matched by
the Schuster solution even in the tail of the cloud. Since the
elongated head-tail structure of L1251 is far from the spherical
symmetry the good fit of the linewidths in the tail makes it
reasonable to assume that the present cloud structure has been
formed by isothermal contraction.}

\maketitle

\section{Introduction}

The \object{Cepheus Flare} is an extended complex of molecular
clouds, where luminous stars are clearly absent, and it harbors
dark clouds forming low mass stars  (for  details see the
comprehensive study of \cite{1998ApJS..115...59K} on this region).
\object{L1251} ($\alpha_{2000}=22^h 36^m \!\! .1;
\delta_{2000}=+75 \degr 16 \arcmin$) is a dark cloud in this
region at a distance of $ 300 \pm 50 pc$
\citep{1993A&A...272..235K} and apparently belongs to this
complex. Already at the advent of molecular radio astronomical
studies in the late sixties the cloud was detected among the ten
brightest OH emission sources on the sky by
\cite{1969ApJ...155L..21C}. The cloud was also listed among
sources of strong formaldehyde emission
\citep{1973ApJ...183..449D,1975A&A....39..435S}.

After these early successes of molecular radio observations L1251
apparently escaped the attention of radio observers. Although
\cite{1982Ap.....18...37K}  discovered a number of $H \alpha$
emission stars associated with L1251, only appeared again in
molecular radio studies in the late  eighties
\citep{1989ApJ...343..773S,1989ApJ...346..168Z,
1989ApJS...71...89B}. \citet{1989ApJ...346..168Z} recognized an
ammonia core in the dense part of the cloud and comparing the
large number of $H \alpha$ objects associated with the cloud,
\citet{1989ApJ...343..773S}  concluded that the star formation
efficiency is anomalously high in L1251. From all of these facts
it was obvious that the star formation processes in this region
needed further detailed investigations. This motivated
\cite{1993A&A...272..235K}  to study the properties and
distribution of faint IRAS point sources along with the $H \alpha$
emission objects. Their study indicated that L1251 has been
forming low-mass stars with an efficiency higher than usually
encountered in dark clouds. The eastern, head region of the cloud
has been found to contain more evolved YSOs than the western
(tail) side.

A census of dense cores carried out by \citet{1996A&A...311..981T}
based on the $NH_3$ 1.3 cm line using the 100 m dish at Effelsberg
resulted in detection of eight ammonia cores with typical size
 of $FWHM \approx 2 \arcmin$ (0.2pc at 350pc distance). Five of
the cores were found to be gravitationally bound. L1251 was
recently surveyed in several mm lines, the structure and
kinematics of the cores were studied e.g. by
\cite{2002ApJ...572..238C}, \cite{1999ApJ...526..788L},
\cite{2001ApJS..136..703L} and \citet{2003A&A...409..941N}.
  Both infall and outward motions were detected in the
gravitationally unstable cores.

\cite{1995Ap&SS.233..175T}  tried to explain the shape of the
cloud, as seen in the CO observations \citep{1993ApJ...406..528G,
1994ApJ...435..279S}, by hydrodynamical modelling assuming an
encounter of a dense molecular cloud with an external shock,
probably originating from a nearby supernova explosion. The
reality of this assumption is supported by the detection of a soft
X-ray excess region east of the cloud by
\cite{1989ApJ...347..231G}. Further confirmation of an SN
explosion comes from the space motion of the runaway star HD203854
\citep{2000MNRAS.319..777K}.

Although the dense gas component of the cloud is well studied,
much less is known about the distribution and properties of
interstellar dust in and around L1251. Recently
\citet{2003AJ....126.1888K} studied the extinction of the dust
component  using {\it B, V, R, I} star counts and found high $R_V$
values indicating grain growth in the head of the cloud.

 In the present paper we
study the spatial distribution of dust, the mass, and the basic
physical properties in the cloud by means of new optical
extinction maps in {\it B, V, R, I} colors. In a subsequent paper
the thermal emission of the dust grains in the dense regions of
the cloud will be analyzed on the basis of far-infrared (120 and
200$\mu m$) maps obtained from the Infrared Space Observatory
(T\'oth et al. in prep.). The two complementary studies will
clarify the spatial distribution of dust with respect to the
distribution of the dense gas, and indicate how the  star
formation activity and the encounter with the external shock has
modified the properties of the dust particles. To obtain direct
information on the spatial distribution of dusty material the
study of star counts is still one of the most reliable approaches.
The basic aim of this  paper is to carry out such analysis.

\section{Input data}

To study the surface distribution of the optical extinction we
obtained star counts in {\it B, V, R, I} colors based on
photographic observations with the 60/90/180 cm Schmidt telescope
of the Konkoly Observatory. The plates were digitized with a pixel
size of 20 microns in a $1.5 \times 1.5$ degree field around L1251
using the PDS microdensitometer of the Vienna Observatory in 1991.
We scanned 4, 3, 3, and 2 plates in $B, V, R_J, I_J$ colors,
respectively. The scans were processed with the ROMAPHOT
photometric programme integrated into the MIDAS data analysis
package. The plates were calibrated via CCD observations performed
with the 1.23 m Ritchey-Chretien telescope of the German-Spanish
Observatory, Calar Alto \citep{1992A&A...255..281B}. The limiting
magnitude of the photographic survey was 19.0, 18.5, 17.5, and
16.5 in $B, V, R_J,$ and $I_J$, respectively. Although these
figures are  less than could be obtained from the Digitized Sky
Survey maps (see e.g. the extinction maps of
\cite{1999A&A...345..965C}), the well defined color system, the
much better photometric calibration and four colors, however,
support using our data. The estimated completeness of star counts
was about 1.5 magnitude above the detection limit of the plates.
The final star count maps of the region were obtained by counting
the stars in each color on pixels of $6 \arcmin \times 6 \arcmin$
size and a $2 \arcmin$ mesh of the star count maps was selected in
both directions. (This resolution approximately corresponds to
those of the 100 $\mu m$ IRAS maps). \cite{1982Ap.....18...37K}
and \cite{1993A&A...272..235K} lists 12 $H\alpha$ objects
(candidate pre-main sequence stars) apparently associated with the
cloud. Their effect on the $R$ and $I$ star counts might be
significant. Consequently, we have omitted them from  further
analysis.

\section{Extinction maps from star counts}
Several studies indicated empirically that the surface
distribution of star counts is an excellent tracer of optical
extinction. (e.g. \cite{1978AJ.....83..363D};
\cite{1985A&A...149..273C}). This means that there is a simple
linear relationship between the logarithmic star counts and the
$a_{cl}$ extinction of a dust cloud:

\begin{equation}\label{anm}
a_{cl} = a\times \log(N(m)) + b(m)
\end{equation}

\noindent where $N(m)$ is the cumulative star count up to a given
$m$ limiting magnitude, the $a$ constant and $b(m)$ depend on the
Galactic longitude and latitude; $-b/a=\log(N_0(m))$ measures the
logarithmic star count in an extinction-free region
\citep{1978AJ.....83..363D}.

In the following we calibrate this expression using a multivariate
statistical method, the $k$-means clustering, and a maximum
likelihood procedure. Based on this calibrated relationship we
assign extinction values to each star count pixel yielding an
extinction map of the cloud. The procedure gives as a byproduct
the distance of L1251 which enables us to calculate the mass of
the cloud. Since we derive the extinction of the cloud in
different colors we also discuss the ratio of the selective to
total extinction.

\subsection{Definition of areas of similar extinction}
The {\it B, V, R, I} star count data on the pixels defined above
represent a distribution of points in a four dimensional {\it B,
V, R, I} parameter space. Following the relationships between star
counts and extinction we assumed that the regions of equal
extinction have equal star counts on the maps. In  other words,
 looking for areas of equal extinction means searching for
points of similar coordinates, i.e. lying close to each other in
the four dimensional parameter space, made up of the pixel values
of the star count maps.

 The linear relationship between the logarithmic star counts and
interstellar extinction predicts a one dimensional manifold in the
four dimensional parameter space, stretched by the extinction, and
there is a Poisson noise superimposed on it by the star counts.
Therefore, to convert star counts into optical extinction we had
to divide the whole star count map into regions of equal counts.
According to Equation (\ref{anm}) the logarithms of the star
counts  scale linearly with the extinction; therefore  we used the
logarithmic {\it B, V, R, I} star counts in this procedure.

\begin{figure}
  \includegraphics[width=9cm]{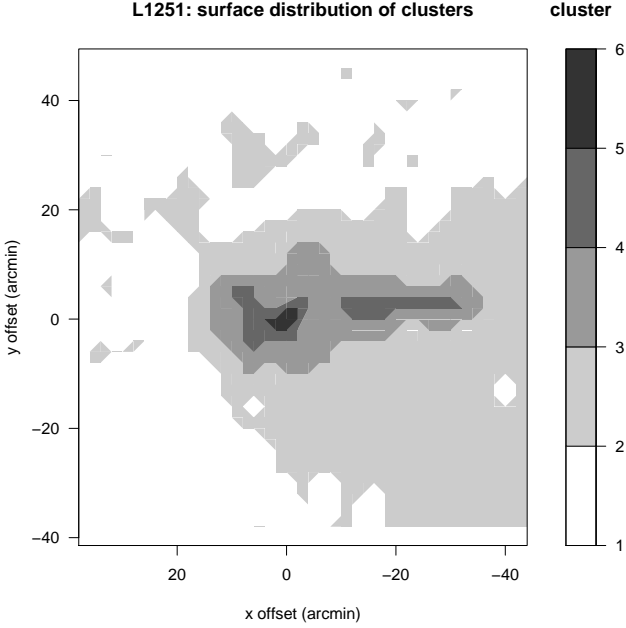}\\
  \caption{Result of the clustering, reprojected onto the
celestial sphere. The areas of same grey level represent pixels in
the map belonging to the same cluster and have similar extinction.
We displayed the cloud in rectangular coordinates in a plane
tangential to the celestial sphere. North is at the top and East
is to the left. The (0,0) position corresponds to
$\alpha_{2000}=22^h 36^m \!\! .1;\, \delta_{2000}=+75 \degr 16
\arcmin$.}\label{gr}
\end{figure}

In order to divide the points into groups of similar extinction in
the parameter space we invoked the technique of $k$-means
clustering (see e.g. \cite{1987mda..book.....M}). K-means
clustering orders the points in the parameter space into $k$
groups. The $k$ number of the groups should be specified before
running the clustering procedure. Assigning the points to any of
the groups proceeds on the basis of some distance measure between
the points. We used squared Euclidean distance.

There are no definite criteria for fixing the value of $k$. By
trial we selected $k=5$ enabling clear separation of the high
extinction regions from those of low extinction and ensuring
enough stars in each cluster for reliable analysis. Figure
\ref{gr} shows the result of the clustering, reprojected onto the
celestial sphere. Areas of same grey level in the map represent
pixels belonging to the same cluster and consequently, having
similar extinction. Table~ \ref{stc} summarizes the number of
stars in each subregion (cluster) in each color. One can infer
from the data of this table that the 5th cluster is scarcely
populated, therefore we excluded it from  further analysis.

\begin{table}
\caption[]{Summary of the star count analysis of L1251,
                number of stars in each cluster and in each color}
\begin{tabular}{ccrrrr}
\hline
$FIELD$ &  $AREA (SqD)$ &  $N_B$ &    $N_V$  &   $N_R$  &   $N_I$ \\
\hline
  1.  &    0.9880   &3982 &  3929  & 3939  & 2723 \\
  2.  &    0.8770   &1883 &  2045  & 2186  & 1530 \\
  3.  &    0.1515   & 101 &   143  &  146  &  138 \\
  4.  &    0.0605   &   7 &    13  &   14  &   15 \\
  5.  &    0.0090   &   1 &     2  &    4  &    4 \\
\hline
\end{tabular}\label{stc}
\end{table}

The central, densest part of L1251 is clearly separated from the
surrounding lower extinction region. It resembles  a  bullet
moving at supersonic speed with respect to the ambient medium. The
less dense area surrounding the bullet-like main body of the cloud
has a form of a bow shock. Accepting this view one can estimate
the  speed of the cloud relative to the ambient medium. The
interaction of the cloud with its surroundings is probably the key
issue in understanding the history of star formation. We will
return to this problem  in Sect. \ref{dis}.

\subsection{Modelling the effect of extinction on the star counts}
\subsubsection{Modelling differential star counts}
To calibrate  star counts in terms of optical extinction we
attempted to model the observed apparent magnitude distribution of
the stars. The starting point of our approach was the Galactic
model of \cite{1992ApJS...83..111W}. Following the basic ideas of
this model we assumed that the spatial distribution of the stars
in the region investigated can be satisfactorily described by
superposition of exponential disks, corresponding to different
types  of stars, and a spheroidal component. The exponential disks
were defined by their scale heights and local stellar densities
near the Sun. We used the data of \cite{1992ApJS...83..111W}  for
the characteristic values of the exponential disks in the model.

To model the effect of the absorbing cloud on the star counts we
assumed that besides the obscuring matter associated with L1251
there is no other significant dust cloud in the line of sight.
This assumption  is quite reasonable due to the high ($+15 \degr$)
galactic latitude of L1251.  As in the \cite{1992ApJS...83..111W}
model we assumed that the diffuse component of interstellar
extinction has the form

\begin{equation} \label{adif}
   a_{diff}(r) =a_0 \sec(b) (1-\exp(-r \sin(b)/h))
\end{equation}

\noindent where $a_0$ is a constant depending on the color
selected, $b$ is the Galactic latitude of the cloud and $h$ is the
scale height of the obscuring material. We added to the extinction
described above a further component in the form of a step
function.

\begin{equation} \label{acl}
a(r) =a_{diff}(r)+ a_{cl}(r) ; \quad a_{cl}(r) = \cases{0, & if $r
< r_{cl}$ \cr  a_{cl}, & if $r > r_{cl}$}
\end{equation}

\noindent The distance of the cloud, $r_{cl}$, and the cloud
extinction, $a_{cl}$, are constants to be estimated within a
cluster obtained by the k-means clustering in the star count
parameter space.

 \subsubsection{Maximum likelihood estimation of the distance and
extinction}

The model described in the previous section allows us to get the
distance and extinction of the cloud using the maximum likelihood
(ML) estimation. According to our model assumption the probability
density of the apparent magnitude of stars in our observed sample
is given by

\begin{eqnarray} \label{am}
& A(m|a_{cl},r_{cl}) = \\
& \omega \sum_{sp} \int_{0}^{\infty} D_{sp}(r) \Phi
(m-5\log(r)-5-a(r|a_{cl},r_{cl})) r^2 dr \nonumber
\end{eqnarray}

\noindent In the above formula $\omega$ is a suitably chosen
normalizing constant, $D_{sp}(r)$ is  spatial density, $\Phi$ is
the luminosity function assumed to have a Gaussian form and $sp$
runs over the spectral types represented in the
\cite{1992ApJS...83..111W}  model. We used a Monte Carlo
simulation  to get a distribution of apparent magnitudes
corresponding to the probability density of $A(m|a_{cl},r_{cl})$.
We used the MC simulated data to calculate the numerical values of
$A$ for the ML procedure.

The likelihood function in our case can be written as

\begin{equation} \label{lf}
  L(a_{cl},r_{cl}) = \sum_{i=1}^{n} \log(A(m_i|a_{cl},r_{cl}))
\end{equation}

\noindent  where $m_i$-s are the observed apparent magnitudes in
one of the colors in our sample. Maximization of
$L(a_{cl},r_{cl})$ with respect to $a_{cl}$ and $r_{cl}$ yields
the ML estimation of the extinction and the distance of the cloud.

\begin{table*}
\caption[]{Summary of the star count analysis of L1251,
                       extinction}
\begin{tabular}{cccccccccc}
\hline
$FIELD$ & $ AREA (SqD)$ & $a_B$  & $\sigma_{a_B}$ & $a_V$ & $\sigma_{a_V}$ & $a_R$ &  $\sigma_{a_R}$ &  $a_I$ &  $\sigma_{a_I}$ \\
\hline
  1.  &    0.9980    & 0.45&  0.05&  0.30&  0.05 & 0.12&  0.05 & 0.10&  0.03 \\
  2.  &    0.8770    & 1.55&  0.10&  1.25&  0.07 & 0.85&  0.07 & 0.63&  0.07 \\
  3.  &    0.1515    & 3.30&  0.15&  2.65&  0.15 & 2.20&  0.15 & 1.75&  0.2  \\
  4.  &    0.0605    & 6.5 &  0.80&  5.2 &  0.80 & $>5.3$&    -  & 4.5 &  0.4  \\
\hline
\end{tabular} \label{abs}
\end{table*}

Performing the ML estimation within all groups given by the
$k$-means clustering and in all colors separately, we obtained the
results summarized in Tables \ref{abs} and \ref{dist}. The results
summarized in Table \ref{abs} enable us to convert the star counts
in different colors into extinction. We return to this calibration
in the following subsection.

\begin{table*}
\caption[]{Summary of the star count analysis of L1251, distance
moduli; the weighted mean of the data gives  $7.58 \pm 0.2$}
\begin{tabular}{cccccccccc}
\hline
$FIELD$ & $AREA (SqD)$ & $r_B$ & $\sigma_{r_B}$ & $r_V$ & $\sigma_{r_V}$ & $r_R$ & $\sigma_{r_R}$ & $r_I$ & $\sigma_{r_I}$ \\
\hline

  1.  &    0.9880   & 7.45 & 0.60 & $>$8.50 &  -  &  -  &   -   &   -  &   -  \\
  2.  &    0.8770   & 7.50 & 0.35 & 7.75  &0.55 & $<$7.6&   -   & $<$7.4 &   -  \\
  3.  &    0.1515   & 7.75 & 0.30 & 7.65 & 0.45 & 7.65&  0.45 & $<$7.4 &   -  \\
  4.  &    0.0605   & 7.05 & 0.7  &$<$7.4   &  -  & $<$7.4&   -   & $<$7.3 &   -  \\
\hline
\end{tabular} \label{dist}
\end{table*}

One may use the calculated distance moduli in Table \ref{dist} to
get an estimate for the distance of L1251. We computed a weighted
mean of the data in the table using weights inversely proportional
to $\sigma$. It resulted in a distance modulus of $7.58 \pm 0.2$
corresponding to $330 \pm 30$ pc.

\subsubsection{Confidence interval for the parameters estimated}
The ML estimation provides a straightforward way to obtain the
confidence interval for the estimated parameters, the extinction
and distance. Denoting the value of the parameters maximizing the
likelihood function with $a_{cl}^{max}$, $r_{cl}^{max}$ and with
$a_{cl}^{true}$ , $r_{cl}^{true}$ their true values we have
asymptotically,  if the sample size goes to infinity,

\begin{equation}\label{ll0}
2 [L(a_{cl}^{max},r_{cl}^{max})-L(a_{cl}^{true},r_{cl}^{true})] =
\chi^2_k  ;\quad  k = 2
\end{equation}

In general, k equals the number of parameters estimated and
$\chi^2_k$ is a $\chi$-square variable with k degrees of freedom
(for the proof of this theorem see \cite{Kendall76}). The
probability that the true values of the parameters are within a
certain region in the k dimensional parameter space is given by

\begin{equation} \label{pdel}
P(\chi^2_k\leq {\chi_0^2}_k)=1-\delta
\end{equation}

The projection of this k dimensional domain which is given by the
$\chi^2_k\leq {\chi_0^2}_k$ inequality yields the confidence
interval of the individual parameter values estimated by the ML
procedure. The ${\chi_0^2}_k, \delta$ pairs are tabulated in the
literature (see e.g \cite{Kendall76}).  The confidence intervals
corresponding to the $1\sigma$ levels are given in the $\sigma$
columns of the tables. In the case of $Field$ 4, due to the small
numbers of stars the confidence interval is not closed towards
higher extinctions and lower distances.

\subsection{Star count - extinction conversion}
\subsubsection{Verifying the linear $\log N(m)$ - extinction relationship}
The results yielded by the ML analysis enabled us to verify the
star count - extinction relationship in each color in our study.
Assuming the functional form of the formula given by Eq.
(\ref{anm}) one can get its constants  by a linear least squares
fitting of the star counts versus extinction given in Table
\ref{abs}. We summarized the results of the least squares fitting
in Table \ref{lsq}.

\begin{table}
\caption[]{Linear least squares fitting of the $a_{cl} = a\times
\log (N(m)) + b(m)$ extinction - star count relation}
\begin{tabular}{ccccc}
\hline
 $color$ & $a$ & $\sigma_a$ & $b(m)$ & $\sigma_b$ \\
 \hline
 $B$     &  $-$4.08   & 0.23  & 5.78 & 0.21 \\
 $V$     &  $-$4.08   & 0.16  & 5.67 & 0.15 \\
 $R$     &  $-$4.39   & 0.30  & 5.86 & 0.28 \\
 $I$     &  $-$4.52   & 0.18  & 5.31 & 0.16 \\
 \hline
\end{tabular} \label{lsq}
\end{table}
Table \ref{lsq} clearly demonstrates that the linear relationship
fits  the data obtained from the ML analysis. This means that the
postulated linearity was convincingly recovered from the ML
analysis performed.

In the $B$ and $V$ colors the slope of the  relationship is the
same while in $R$ and in $I$, in particular, it significantly
differs. In the literature the inverse value of $a$ is usually
given. Using the value of $a$ obtained for $B$  would give a bias
of about $0.44\,\log [N_0(m)/N(m)]$ $mag$ in the estimation of the
optical extinction, in the $I$ color.

\subsection{Surface distribution of the obscuring material}
\begin{figure*}
  \includegraphics[width=8cm]{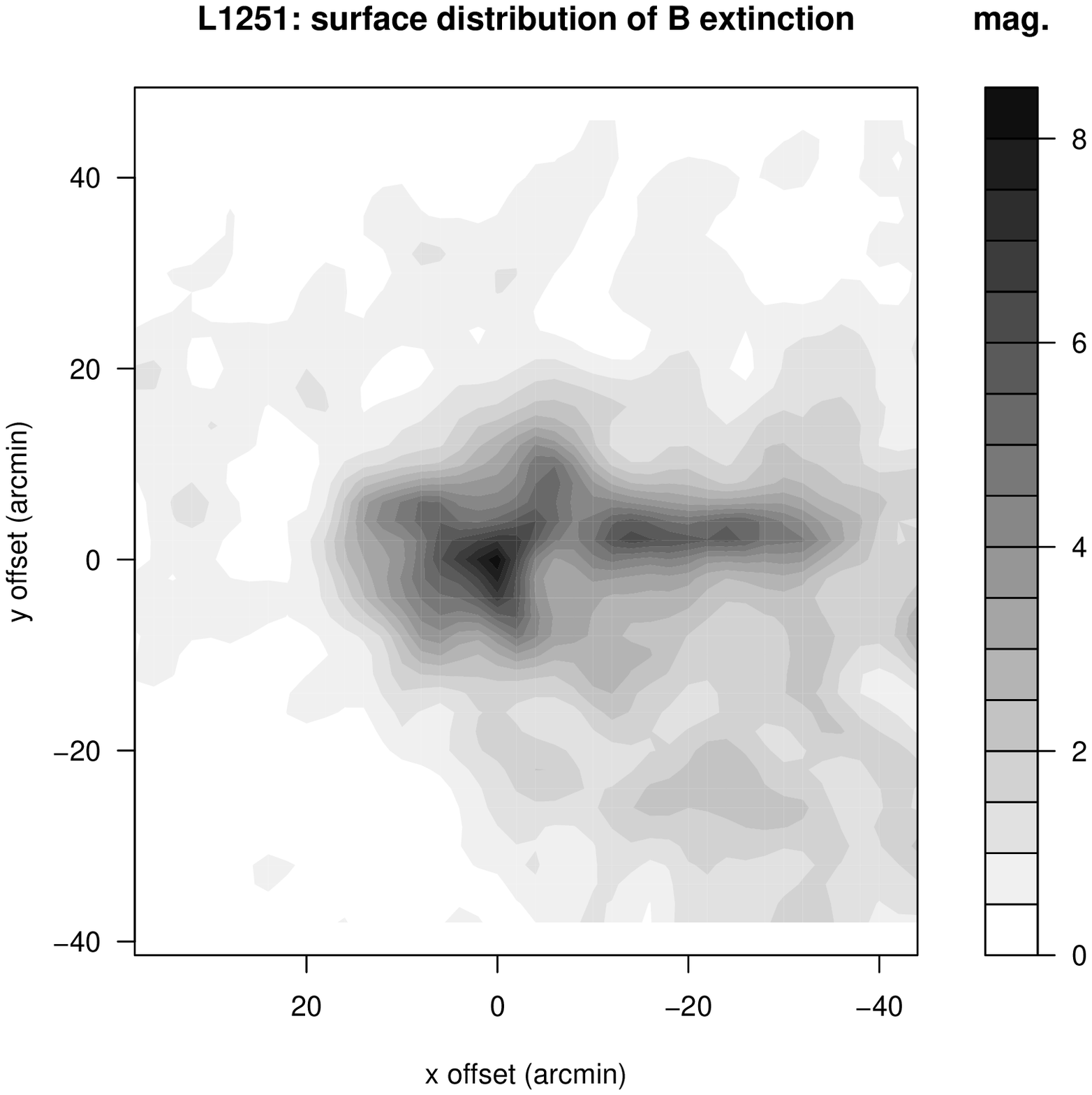}\includegraphics[width=8cm]{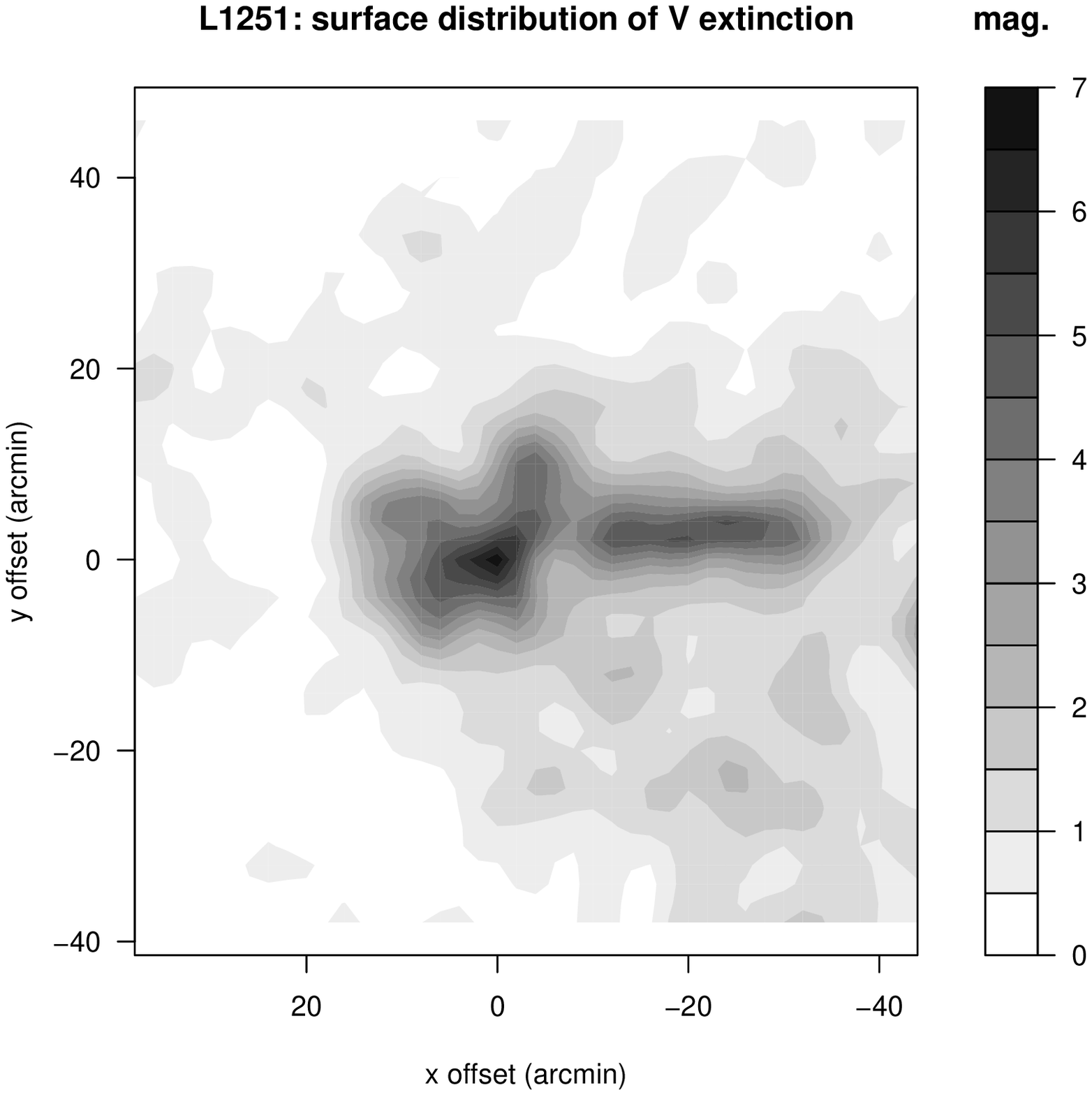}\\
  \includegraphics[width=8cm]{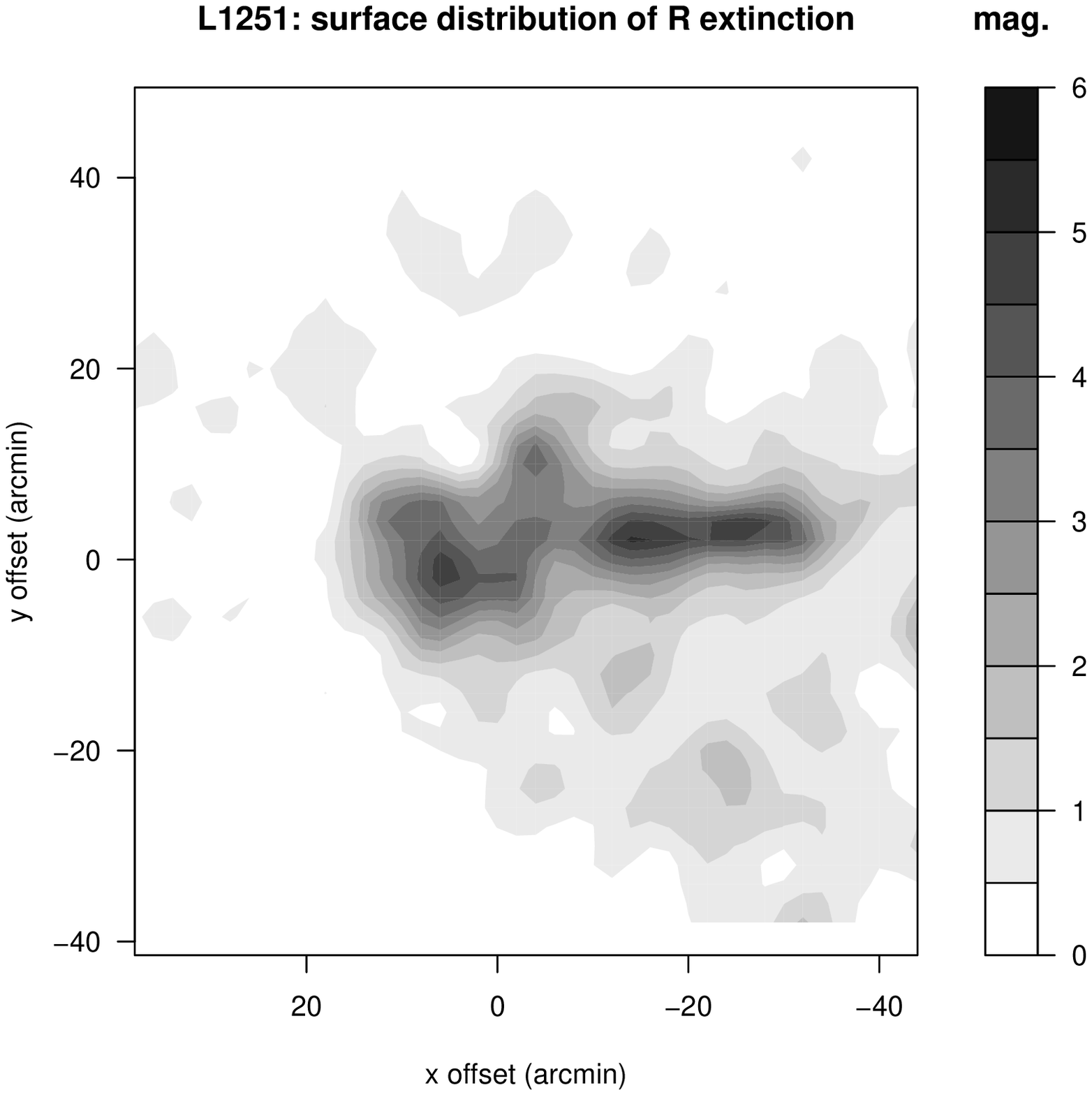} \includegraphics[width=8cm]{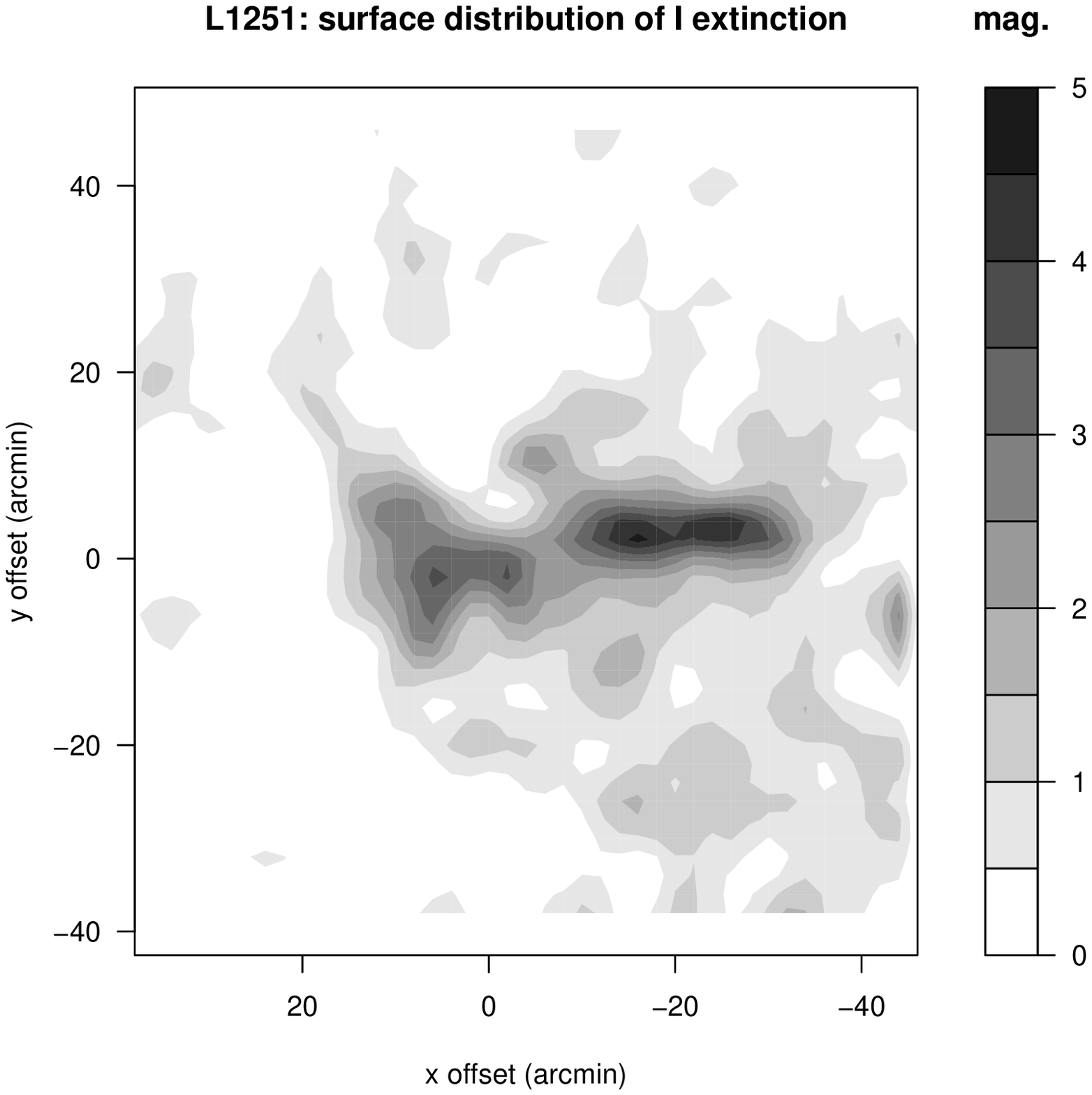}
  \caption{Contour maps of the extinction in the $B,V,R$ and $I$
color. The `flying bullet' form of the main body and the bow shock
displayed in Fig. \ref{gr} is clearly visible. The coordinates of
the (0,0) position are same as in Fig. 1.}\label{b}
\end{figure*}

Based on the calibration procedure one may assign extinction to
each pixel in the  star count maps, in all of the four colors
studied. Fig. \ref{b}  shows the contour maps of the extinction
obtained in this way. The main body of the cloud stands out with a
contrast of several magnitudes from the less obscured region
behind the bow shock.

The overall distribution of the obscuring material, obtained from
our study and that of \citet{2003AJ....126.1888K},  has a
reasonably good correlation. There are, however, remarkable
differences between them, in particular in the densest part of the
cloud. The probable reason for these discrepancies lies in the
different ways used in obtaining extinction maps from the surface
distribution of the stars in the region of L1251. Both studies had
a mesh of 2' resolution but applied different kind of smoothing.
We used a boxcar of 6'x 6' size while the other study was a
Gaussian filtering of 6' resolution. It is probably important to
note that the tails of the Gaussian filter give contributions from
much wider areas, in particular in the densest part where the star
counts have low values.

A further reason for the discrepancy might originate from the
method of converting the star count maps into extinction. Our
method accounts for the foreground stars whose contribution
increases the counts and decreases correspondingly the estimated
extinction values. We made simulations of the
\cite{1992ApJS...83..111W} model assuming a limiting magnitude of
about 19 mag and 6 mag of extinction at the distance of L1251. The
conventional method of  star count extinction conversion resulted
only in a 5 mag. value, i.e. one magnitude less than the true (6
mag.) extinction.

\section{Discussion} \label{dis}
\subsection{Shape of the cloud} \label{shape}
We have already indicated above that the extinction map derived
from the star counts resembles a body flying  at hypersonic speed
across an ambient medium.

The shape of L1251 was accounted by \cite{1995Ap&SS.233..175T}
 for a shock wave passing the cloud and
produced by a nearby supernova. They showed that the cooling by
the $H_2$ molecules plays an important role in the formation of
the observable shape of the cloud.

The presence of a bow shock as indicated by the extinction maps
suggests another type of  cloud-environment interactions. The
blunt body solutions of  hypersonic flows are standard topics in
 textbooks (see e.g. \cite{Hayes59}. For the astrophysical
context of the problem see the works of \cite{1985A&A...147..220R}
and \cite{1998MNRAS.297..383C}). One can identify the tail of
L1251 with a wake of the head of the cloud, typical of  blunt body
hypersonic streaming patterns.

In the following we do not try to fit the  form of the bow shock
using a solution of the blunt-body problem. This solution would
allow a better understanding of the role of different significant
physical parameters defining a particular fit, however this is
beyond the scope of the present paper. An obvious significant
parameter would be the Mach number of the flow which is easy to
calculate from the angle between the two asymptotes of the bow
shock \citep{Hayes59}. The Mach number is given by
$M=\sin(\alpha/2)^{-1}$ where $\alpha$ is the angle between the
asymptotes. It yielded $M\approx2$ in our case.

 The symmetry axis of the
bow shock has a tilt of about 10 deg to the main axis of L1251.
 The symmetry axis of the
bow shock points toward the center of the bubble discovered by
\cite{1989ApJ...347..231G}. \cite{2000MNRAS.319..777K} found a
runaway star which also might have been caused by the SN explosion
that was probably responsible for the bubble and derived an age of
$10^6$ yrs. According to this picture the bow shock resulted from
an encounter of the cloud with the wind coming from the interior
of the bubble.

\subsection{Properties of the obscuring material}

\begin{figure*}
  \includegraphics[width=8.5cm]{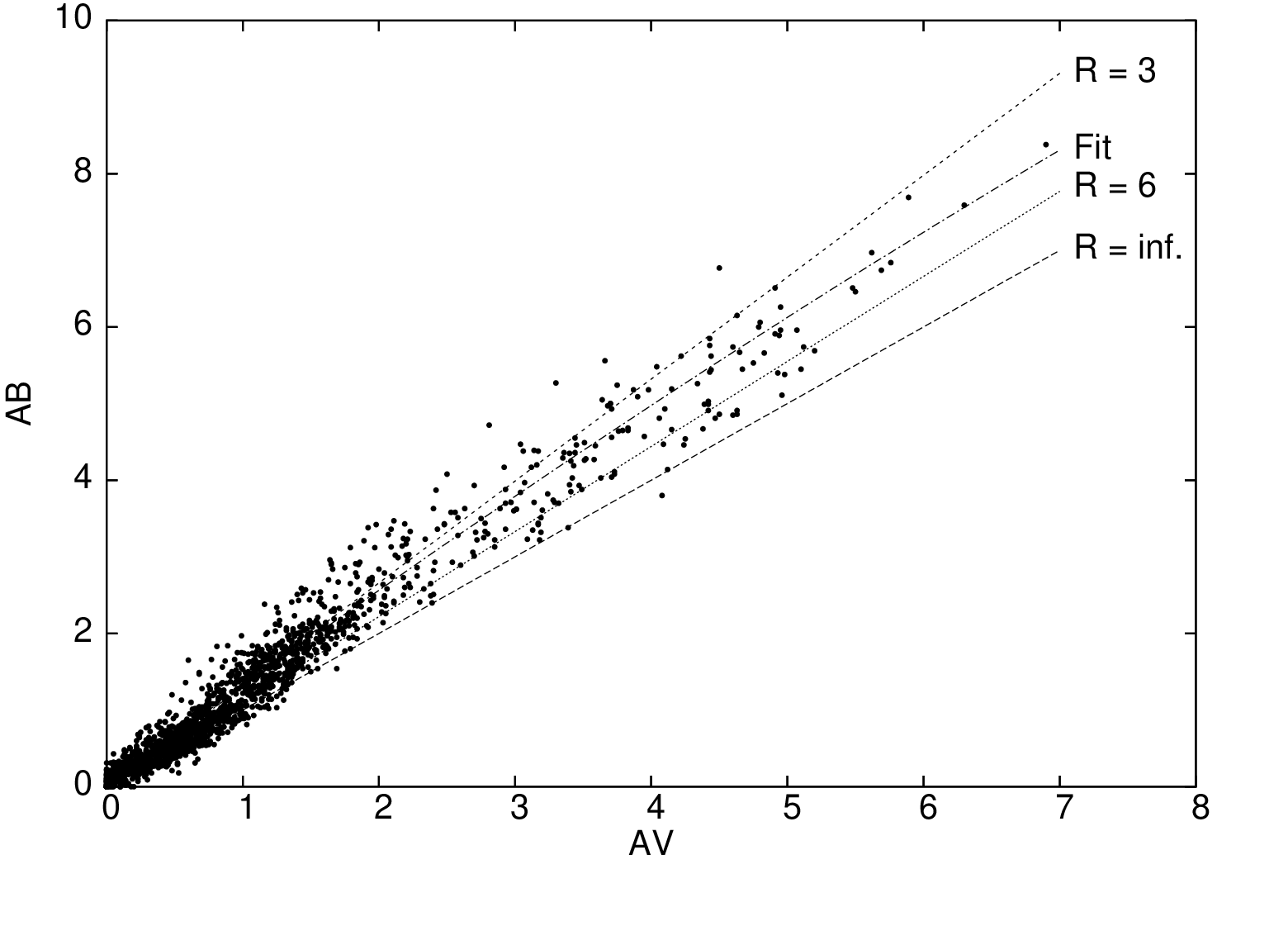}
  \includegraphics[width=7.1cm]{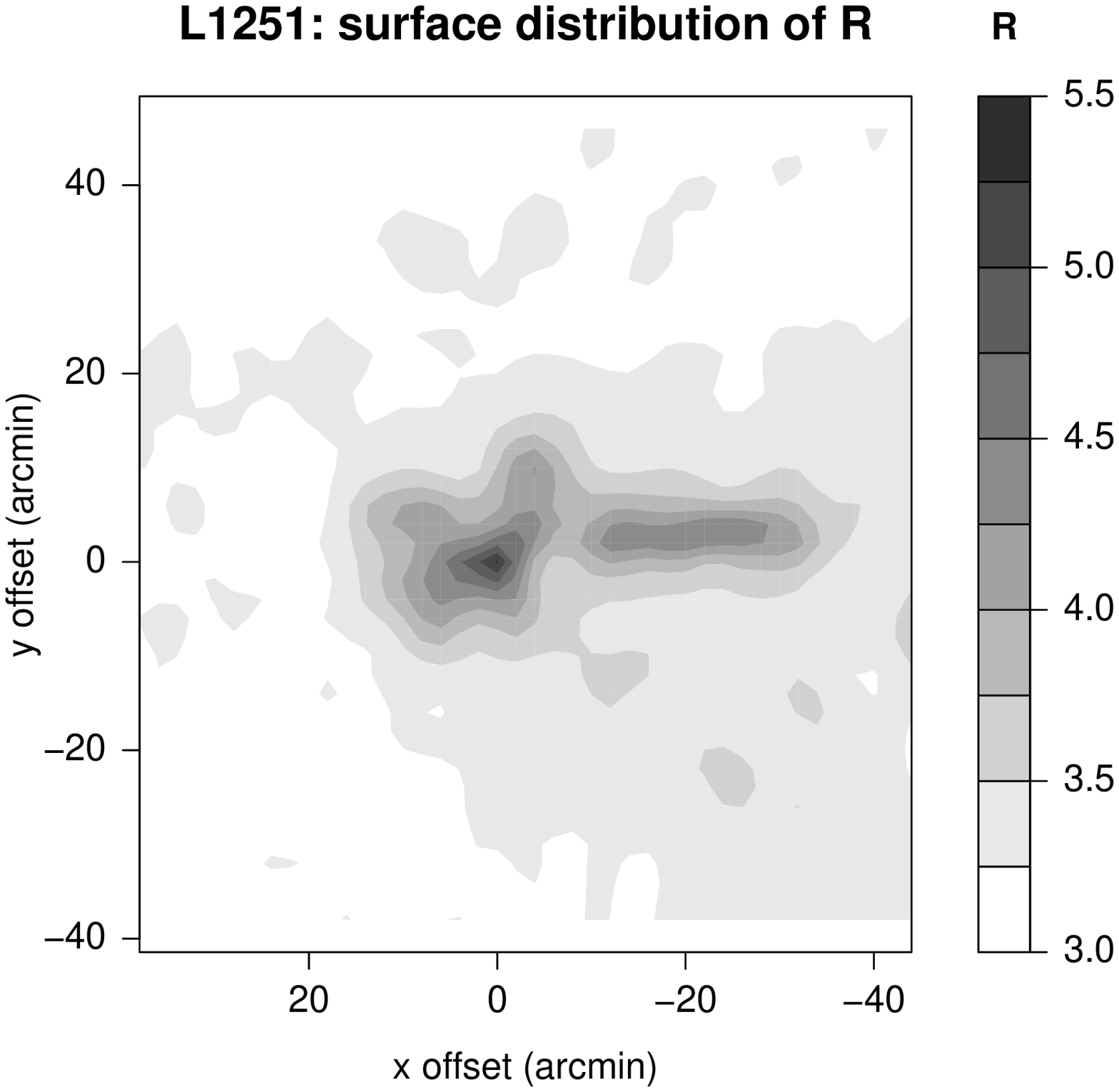}\\
  \caption{Scatterplot between $a_V$ and $a_B$ (left panel). The
relationship between $a_V$ and $a_B$ can be written in the form of
$a_B=(1+1/R) \times a_V$. We marked with lines (labelled with the
corresponding $R$ values) the relationships between $a_V$ and
$a_B$ assuming that $R=3$ (close to the canonical value for the
general interstellar matter) and $R=6$. The line assuming that the
interstellar extinction does not depend on the color (grey
approximation) is also marked with $R=inf$. Assigning $R_V$ to
each pixel we obtained a map of the total to selective extinction
(right panel). One may infer from this map that the high values of
$R_V$ are concentrated only in the densest parts of L1251. The
coordinates of the (0,0) position are same as in Fig.
1.}\label{avab}
\end{figure*}

The estimation of the total extinction in different colors makes
it possible to get the value of the total to selective extinction
$R_V=a_V/E_{B-V}$ . The scatterplot between $a_V$ and $a_B$ is
displayed in Fig. \ref{avab}. The relationship between $a_V$ and
$a_B$ can be written in the form of $a_B=(1+1/R_V) \times a_V$. We
marked with lines (labelled with the corresponding $R_V$ values)
in Fig. 3 the relationships between $a_V$ and $a_B$ assuming that
$R_V=3$ (close to the canonical value for the general interstellar
matter) and $R_V=6$. The line assuming that the interstellar
absorption does not depend on the color (grey approximation) is
also marked with $R_V=inf$.

Figure \ref{avab} clearly demonstrates that above  $a_V=3$ (i.e.
within the main central body of L1251) the points  depart from the
$R_V=3$ line and approach the lines of higher $R_V$ as one moves
to higher extinction. Departure from the canonical $R_V=3.1$
towards higher values is a common behavior of dense clouds  (see
the review by \cite{1990ARA&A..28...37M}).

Postulating a relationship in the form of $a_B= \alpha \times
a_V$, where $\alpha=(1+1/R_V)$, and substituting $R_V$ values
between 3-6, we can infer that $\alpha$ depends only weakly on
$a_V$ and may have approximately a form of
$\alpha(a_V)=\alpha_0+\alpha_1 \times a_V$. Assuming that
$R_V=1+1/\alpha_0=3.1$ we made a least squares fitting to get
$\alpha_1$. In this way we obtained $R_V(a_V)=1/(0.32-0.019\times
a_V)$ and $a_B=\alpha(a_V)\times a_V$ gave a reasonably good fit
to the points in Fig. \ref{avab}.


Based on this result we may assign a value of  $R_V$ to each pixel
in the $a_V$ extinction map of L1251.  We obtained a map of the
total to selective extinction as given in Fig. \ref{avab}. One may
infer from this map that the high values of $R_V$ are concentrated
only in the densest parts of L1251.

The canonical dependence of the interstellar extinction on the
wavelength is realized by the standard relation of
\cite{1979ARA&A..17...73S} which can be well represented by a
linear function of the extinction on $1/\lambda$ in the range of
the {\it B, V, R, I} colors used in our study.

Departure from the standard $R_V=3.1$ value above $a_V=3$ in dense
interstellar clouds, as we found in the case of L1251, was also
obtained by \citet{2001ApJ...547..872W} in the Taurus region. They
claimed that for extinctions $a_V > 3$ real changes in grain
properties may occur, characterized by observed $R_V$ values in
the range of 3.5-4.0. A simple model for the development of $R_V$
with $a_V$ suggested that $R_V$ may approach values of 4.5 or more
in the densest regions of the cloud. According to
\citet{2001ApJ...547..872W} the transition between ``normal"
 and ``dense cloud" extinction occurs at $a_V \approx
3.2$, a value coincident with the threshold extinction above which
$H_2O$-ice is detected on grains within the cloud.

The $R_V$ values derived in our analysis correspond to those
obtained by \citet{2003AJ....126.1888K} within the limits of the
statistical errors. There are differences, however, in the surface
maps between the two studies. To obtain $R_V$ values we used the
relationship between the $a_B$ and $a_V$ values so our surface map
reflects  that of the extinction in the $V$ color. On the
contrary, \citet{2003AJ....126.1888K} applied an adaptive
averaging of $B-V$ and $V-I$ color indices by varying the size of
the smoothing window, keeping constant the number of stars in it
but at the cost of the spatial resolution, in particular in the
densest part of the cloud.

\subsection{Mass of the cloud}
Knowing the distance of L1251 we converted the $a_V$ extinction
values into the mass using the empirical formula given by
\cite{1978AJ.....83..363D}. According to this formula

\begin{equation} \label{mas}
M=(\alpha d)^2 \mu {N_H \over a_V} \sum_i a_V(i)
\end{equation}

\noindent where $M, \alpha, d, \mu, a_V(i)$ are the mass of the
cloud, the angular size of a pixel, the distance, the mean
molecular mass and the extinction of a pixel, respectively;
$N_H/a_V=1.87 \times 10^{21} cm^{-2}mag^{-1}$ and $N_H=N_{HI} + 2
N_{H_2}$. Based on  Dickman formula we computed the mass of the
different subregions of the cloud as given in Table \ref{dick}.

\begin{table}
\caption[]{Mass of the cloud (using the formula of
\citet{1978AJ.....83..363D})}
\begin{tabular}{ccrr}
\hline
FIELD &  AREA (SQD) &  MASS ($M_{\odot}$) &  NO. OF PIX.\\
\hline

  1.  &    0.9880 &  153.34 &    887  \ \ \ \ \\
  2.  &    0.8770 &  490.48 &    784  \ \ \ \  \\
  3.  &    0.1515 &  212.60 &    136 \ \ \ \ \\
  4.  &    0.0601 &  132.05 &     54  \ \ \ \  \\
  5.  &    0.0089 &   26.07 &      8  \ \ \ \  \\
\hline
\end{tabular} \label{dick}
\end{table}

Fields 3, 4 and 5 represent the main body of L1251. Based on Table
\ref{dick} the total mass of this part of the cloud is 371
$M_{\odot}$. This value can be compared with the figure obtained
by  \cite{1994ApJ...435..279S} based on $C^{18}O$ measurements,
taking into account that the field occupied by our main body lies
completely inside the region covered by the $C^{18}O$ study and
contains only 85 \% of the mass calculated from it. Re-scaling
this fraction of mass to the distance we obtained in this paper
one gets 422$M_{\odot}$, surprisingly close to our estimated
value. Studying a somewhat larger area than ours,
\citet{1994JKAS...27..159L} obtained $610 \, M_\odot$ from his
$^{12}CO$ and $^{13}CO$ measurements.

\subsection{Mass model of the head of L1251}
  In the following we try to model the
head of the cloud assuming spherical symmetry.

\begin{figure}
  \includegraphics[width=6.2cm, angle=270]{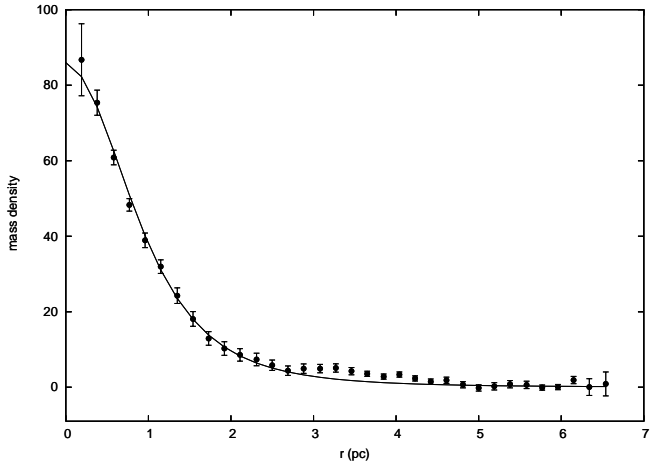}\\
  \caption{Fitting the radial distribution of the surface
mass density by a Schuster sphere. Up to 2.5 pc distance from the
center the fit is very good. Beyond this distance, however, there
is an excess of the observed mass density due to the tail of the
cloud.}\label{mass}
\end{figure}

Using the  data obtained with  VLT of the ESO
\cite{2001Natur.409..159A} modelled the spatial structure of the
Bok globule Barnard 68. They supposed the globule to be an
isothermal Bonnor-Ebert sphere
\citep{1956MNRAS.116..351B,1955ZA.....37..217E} in  pressure
equilibrium with the external, much hotter and less dense
interstellar medium.

We have not required isothermality in our analysis. The isothermal
solution is a specific case of the class of polytropic spheres
studied extensively by \citet{Emden}. The isothermal solution
represents the case when the polytropic index $n=\infty$.
Actually, the mass of an isothermal sphere is infinite and the
limit between the finite and infinite mass solutions is $n=5$.
Although the mass is finite in this case the size of the sphere is
infinite. One obtains finite size solutions only in the $n<5$
case. Recently, \citet{2001ApJ...555..863M} studied the properties
of polytropic spheres near $n=5$ and found them very suitable to
characterize the structure of molecular cloud cores.

We fixed the mass center of the sphere at the maximum value of
$a_V$. This choice is quite reasonable in view of Equation
(\ref{mas}).  We then averaged the $a_V$ values over concentric
annuli. This procedure gave a radial profile of the extinction of
the head of L1251. This profile can be converted into the surface
mass density by Equation (\ref{mas}).

Evaluating the radial profile obtained from the observed data we
projected the $n=5$ polytropic sphere \citep{Schuster}  to get its
surface mass density distribution. The Schuster sphere has two
free parameters to be adjusted: the central mass density and a
scale parameter.  We get the fit displayed in Fig. \ref{mass}. Up
to 2.5pc distance from the center the fit is very good. Beyond
this distance, however, there is an excess of the observed mass
density due to the tail of the cloud which significantly distorts
the spherical symmetry of the head.

The density parameter of the fitted Schuster sphere,  $45
M_{\odot} pc^{-3}$ ($3.06 \times 10^{-21} {\rm gcm}^{-3}$), gives
an estimate of the central mass density of the head of L1251.
Comparing the finite size polytropic solutions of $n<5$ with those
of the Schuster sphere (i.e. $n=5$) one can infer that they
concentrate more mass at finite distances and give a much worse
fit to the points observed.

\begin{figure}
  \includegraphics[width=6.2cm,angle=270]{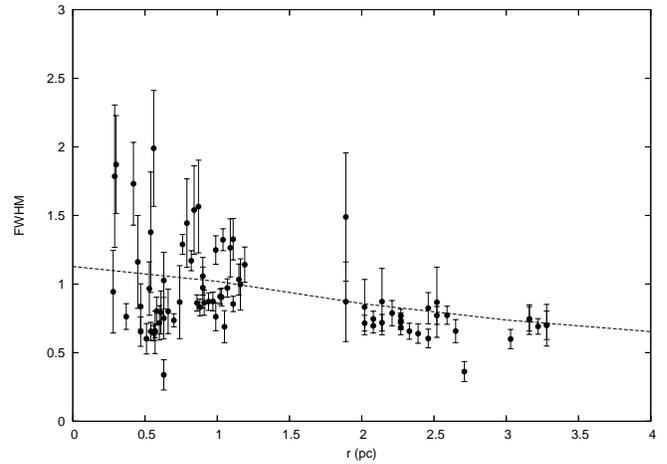}\\
  \caption{$FWHM$ of the projected velocity profile of a
Schuster sphere (dashed line) and the $NH_3$ molecule measured by
\citet{1996A&A...311..981T}. There is a considerable scatter of
the line widths in the head of the cloud around the mean which is
well matched by the Schuster solution. The Schuster curve gives an
unexpectedly good fit in the tail region.}\label{vel}
\end{figure}

Assuming a polytropic gas sphere one can compute the radial
velocity dispersion profile from the fitted density profile, based
on the polytropic equation of state. The $p/\rho$ ratio of the
pressure and mass density gives the $T$ temperature, i.e. the
velocity dispersion. The velocity dispersion projected onto the
celestial sphere can be directly compared with that observed. The
projected $FWHM$ of the V-profile of a Schuster sphere is
displayed in Fig. \ref{vel}, along with those of the $NH_3$
molecule measured by \citet{1996A&A...311..981T}.

There is a considerable scatter of the measured line widths in the
head of the cloud around the mean which is well matched by the
Schuster solution. The Schuster curve, however, gives an
unexpectedly good fit in the tail region. This good fit is not
expected far from the head since due to the tail the mass
distribution  drastically differs from the spherical symmetry.
However, we may assume that our predicted FWHMs of the NH3 line
refers to the initial stage and those measured by
\citet{1996A&A...311..981T} to the final stage of the formation of
density enhancements in the tail region. We conclude therefore
that the formation of structures in the tail left the linewidth
practically unchanged. This indicates an isothermal contraction.
However, the question remains open: what kind of instability
played a significant role in forming the density enhancements in
the tail? We note that according to \citet{1996A&A...311..981T}
the thermal energy is dominant over the turbulent energy in the
NH3 cores of the tail region and thus we cannot exclude the
scenario that thermal instability played a significant role in the
cloud fragmentation.

\section{Summary and conclusions}

In this paper we studied the $B, V, R, I$ star count maps of L1251
 to derive the spatial structure of the obscuring matter
in the cloud. We assumed that the pixel values in the star count
maps form a one dimensional manifold in the $BVRI$ four
dimensional parameter space and defined areas of similar
extinction by means of multivariate $k-means$ clustering. After
defining areas of equal extinction we derived the amount of
obscuration using a maximum likelihood procedure based on a Monte
Carlo simulation of the \cite{1992ApJS...83..111W} model. As a
byproduct of the model fitting we obtained the $330\pm30 \, pc$
distance of the cloud.

The extinction maps clearly showed the main body of the cloud and
a less dense region having a form of a bow shock of a blunt body.
The form of the bow shock allowed us to calculate the approximate
Mach number ($M\approx2$) of the streaming around the head of
L1251.

Comparing the obscuration of the dust in different colors we
calculated the dependence of the total to selective extinction
($R_V$ value) on the visual extinction. The  total to selective
extinction exceeded $R_V=5$ in the densest part of the cloud.

 Using Equation (\ref{mas}) to convert the optical extinction distribution into
surface mass density we obtained the mass of 371 $M_{\odot}$ for
the cloud, in reasonably good agreement with the 410 $M_{\odot}$
 obtained by \citet{1994ApJ...435..279S}, based on $C^{18}O$
measurements.

Assuming a spherical  symmetry for the head of L1251 we computed
the radial distribution of the mass and compared it with a
polytropic model of $n=5$ (Schuster sphere). Up to $r=2.5 \, pc$
from the center of the mass the fit is excellent but beyond this
distance the observed points start to depart remarkably from the
fit due to the drastic distortion of the spherical symmetry by the
tail of L1251.

In the head of L1251 the Schuster model fit matches well the mean
FWHM of the 1.3 cm NH3 line measured by
\citet{1996A&A...311..981T}. It gives an unexpectedly good fit in
the tail region indicating that isothermal contraction played a
significant role in forming the density enhancements.

\acknowledgements{This work was partly supported by OTKA grants
T\,024027, T\,34584 and T\,037508 of the Hungarian Scientific
Research Fund.
  P.\'A. acknowledges the support of the Bolyai Fellowship.}

\end{document}